\documentstyle[12pt]{article}
\begin{document}

\begin{flushright}
{}\hfill
MRI-PHY/P991237
\end{flushright} 

\begin{center}
{\huge\bf Vacuumless cosmic strings in Brans-Dicke theory}\\
\vspace{5mm}
A.A.Sen\footnote{e-mail:anjan@mri.ernet.in}\\
Mehta Research Institute\\
Chhtanag Road, Jhusi\\
Allahabad, 211019\\
India\\
\end{center}
\vspace{10mm}
{\centerline{\bf Abstract}}
The gravitational fields of vacuumless global and gauge strings have been
studied in Brans-Dicke theory under the weak field assumption of the field
equations. It has been shown that both global and gauge string can have
repulsive as well as attractive gravitational effect in Brans-Dicke theory
which is not so  in General Relativity. \vspace{5mm}
\section{Introduction}

Spontaneous symmetry
breaking in the gauge field theories may give rise to some  topologically
trapped regions of a false vacuum, namely domain walls, cosmic strings or
monopoles, depending on the dimension of the region ~\cite{Kibb}. In 
cosmology, these defects have  been put forward as possible source for the density
perturbations which seeded the galaxy formation ~\cite{VS}. 

A typical symmetry breaking lagrangian is of the form
$$
{\cal{L}} = {1\over{2}}\partial_{\mu}\phi^{a}\partial^{\mu}\phi^{a} -V(f),
\eqno{(1.1)}
$$
Where $\phi^{a}$ is a set of scalar fields, $a=1,2,...N$,
$f=(\phi^{a}\phi^{a})^{1/2}$ and $V(f)$ has a minimum at a non zero value
of $f$. The model has $O(N)$ symmetry and domain walls, strings, monopoles
are formed for $a=1,2,3$ respectively. One has to add gauge field in the
above lagrangian and should replace $\partial_{\mu}$ by a gauge covariant
derivative, to study the structure of gauge defects.

It has been recently suggested by Cho and Vilenkin~\cite{CV1} that
topological defects can also be formed in the models where $V(f)$ is
maximum at $f=0$, and it decreases monotonically to zero for
$f\rightarrow\infty$ without having any minima. For example,
$$
V(f) = \lambda M^{4+n}(M^{n}+f^{n})^{-1},
\eqno{(1.2)}
$$
where $M$, $\lambda$ and $n$ are positive constants. This type of
potential
can arise in nonperturbative superstring models~\cite{CV3}. Such potential
having a power law tail for large $\phi$ has also been considered by
authors~\cite{CV4} inorder to reconcile the low dynamical estimates of the
mean mass density with negligibly small scale curvature which is preferred
by inflation. In recent years, potential of this type has been discussed
in
so called "{\it{quintessence}}" models of inflation~\cite{CV5}. Defects
arising in
these models are termed as Vacuumless. In a recent paper, Cho and Vilenkin
have studied the gravitational fields of such vacuumless defects in
General Relativity[GR]~\cite{CV2}

At sufficient high energy scales it seems likely that gravity is not given
by the Einstein's action, but becomes modified by the superstring terms.
In the low energy limit of this string theory, one recovers Einstein's
gravity along with a scalar dilaton field which is non minimally coupled to
the gravity~\cite{AAS18}. On the other hand, scalar tensor theories, such as
Brans-Dicke theory(BD)~\cite{BD}, which is compatible with the Mach's
principle, have been considerably revived in the recent years. It was
shown by La and Steinhardt~\cite{LS} that because of the interaction of
the BD scalar field with the Higgs type sector, the exponential inflation
in Guth's model ~\cite{Guth} could be slowed down to power law one and the
graceful exit in the inflation is thus completed via bubble nucleation.
Although dilaton gravity and BD theory arise from entirely different
motivations, it can be shown that the formar is a special case of the
latter at least formally~\cite{AAS}. As we have mentioned earlier that
these
vacuumless defects may be formed in the supersymmetric phase transition in
the early universe. So it may be relevant to study how these defects
interact with
BD dilaton field which arises in the low energy superstring theories.
Another motivation for studying gravitational properties of defects in BD 
theory  is that
only defects we can hope to observe now are those formed after or near the
end of inflation, and the formation of such superheavy defects is
relatively easy to arrange in Brans-Dicke type theory~\cite{CKL}.

In this work we have studied the gravitational fields of vacuumless
global and gauge strings in BD theory under the weak field approximation
of the field equations. The paper is organised as follows: in section 2 we
have briefly outlined
the work of Cho and Vilenkin for the vacuumless string in GR. In section 3
we have given the solutions of spacetimes for global and gauge vacuumless
string in BD theory under the weak field approximation. The paper ends
with a conclusion in section 4. 

\section{A brief review of vacuumless string in GR}

In this section  we review the earlier work of Cho and
Vilenkin~\cite{CV2}. For global vacuumless string the flat spacetime
solution for $f(r)$ is given by
$$
f(r) = aM(r/\delta)^{2/(n+2)},
\eqno{(2.1)}
$$
where $\delta = \lambda^{-1/2}M^{-1}$ is the core radius of the string;
$r$ is the distance from the string axis and $a =
(n+2)^{2/(n+2)}(n+4)^{-1/(n+2)}$. The solution (2.1) applies for
$$
\delta<<r<<R,
\eqno{(2.1a)}
$$ 
where $R$ is cut off radius determined by the nearest string.

For gauge vacuumless strings, which have magnetic flux localized within a
thin tube inside the core, the scalar field outside the core is given by 
$$
f(r) = aLn(r/\delta) +b.
\eqno{(2.2)}
$$
But here $a$ and $b$ are sensitive to cut off distance $R$:
$$
a \sim M(R/\delta)^{2/(n+2)}[Ln(R/\delta)]^{-(n+1)/(n+2)},
\hspace{3mm} b \sim aLn(R/\delta).
\eqno{(2.3)}
$$
For a vacuumless string the spacetime is static, cylindrically symmetric
and also has a symmetry with respect to Lorentz boost along the string axis.
One can write the
corresponding line element as 
$$
ds^{2} = B(r)(-dt^{2} + dz^{2}) + dr^{2} + C(r)d\theta^{2}.
\eqno{(2.4)}
$$

The general energy momentum tensor for the vacuumless string is given by
$$
T^{t}_{t} = T^{z}_{z} = {f^{'2}\over{2}} +
{f^{2}(1-\alpha)^{2}\over{2C}} + {\alpha^{'2}\over{2e^{2}C}} + V(f)
\eqno{(2.5a)}
$$
$$
T^{r}_{r} = - {f^{'2}\over{2}} +
{f^{2}(1-\alpha)^{2}\over{2C}} - {\alpha^{'2}\over{2e^{2}C}} + V(f)
\eqno{(2.5b)}
$$
$$
T^{\theta}_{\theta} = {f^{'2}\over{2}} -
{f^{2}(1-\alpha)^{2}\over{2C}} - {\alpha^{'2}\over{2e^{2}C}} + V(f)
\eqno{(2.5c)}
$$
Where string ansatz for the gauge field is $A_{\theta}(r) =
-{\alpha(r)\over{er}}$. The $T^{\mu}_{\nu}$'s with $\alpha=0$ are that for
global string.

Under the weak field approximation one can write 
$$
B(r) = 1 + \beta(r), \hspace{2mm} C(r) = r^{2}(1+\gamma(r)),
\eqno{2.5d)}
$$
where $\beta, \gamma << 1$.For global vacuumless string, one can use the
flat space approximation for $f(r)$ in (2.1) for $r>>\delta$ and the form
of $V(f)$ given in (1.2). Then the solution for the spacetime under weak
field approximation is given by~\cite{CV2}
$$
ds^{2} = (1+2\Phi)(-dt^{2} + dz^{2}) + dr^{2} + (1+m\Phi)d\theta^{2}
\eqno{(2.6)}
$$
where $\Phi = -KGM^{2}(r/\delta)^{4/(n+2)}$ ,
$K=\pi(n+2)^{2}/2a^{n}$ and $m$ is an arbitary
constant. The linearized approximation is valid for $\Phi(r)<<1$ and
from (2.1) this is equivalent to $f(r)<<m_{p}$ where $m_{p}=1/\sqrt{G}$ is
the
Planck mass.

 For gauge vacuumless string the energy momentum tensor with
$f(r)$ given in (2.2) can be approximated as~\cite{CV2}
$$
T^{t}_{t} = T^{z}_{z} = T^{\theta}_{\theta} = -T^{r}_{r} =
{f^{'2}\over{2}} = {a^{2}\over{2r^{2}}}.
\eqno{(2.8)}
$$
This form of the energy momentum is valid for 
$$
r<<R/Ln^{1/2}(R/\delta),
\eqno{(2.8a)}
$$
where $R$ is the cut off radius determined by the nearest
string~\cite{CV2}.
The complete solution of the line element under weak field approximation
is given by
$$
ds^{2} = (1+2\Phi)(-dt^{2} + dz^{2}) + dr^{2} +(1-4\Phi)d\theta^{2}
\eqno{(2.9)}
$$
where $\Phi = 2\pi Ga^{2}Ln(r/\delta)$ and $a$ is given by (2.3).
\section{Vacuumless string in BD theory}

The field equation in the BD theory are written in the form
$$
G_{\mu\nu}={T_{\mu\nu}\over{\phi}}+{\omega\over{\phi^{2}}}(\phi_{,\mu}\phi_{,\nu}-{1\over{2}}g_{\mu\nu}\phi_{,\alpha}\phi^{,\alpha})+
{1\over{\phi}}(\phi_{,\mu;\nu}-g_{\mu\nu}\Box\phi),
\eqno{(3.1)}
$$
$$
\Box\phi = {8\pi T\over{2\omega+3}},
\eqno{(3.2)}
$$
Where $\phi$ is the scalar field, $\omega$ is the BD parameter and $T$
denotes the trace of the energy momentum tenosr $T^{\mu}_{\nu}$~\cite{BD}.
In the weak field approximation in BD theory one can assume
$g_{\mu\nu} = \eta_{\mu\nu} + h_{\mu\nu}$ where $|h_{\mu\nu}|<<1$ and
$\phi(r) = \phi_{0}+\epsilon(r)$ with $ |\epsilon/\phi_{0}|<<1$ where $
1/\phi_{0} = G_{0} = {(2\omega_+3)\over{(2\omega+4)}}G.$

It has been shown recently by Barros and Romero~\cite{BR}, that in the
weak field approximation the solutions of the BD equations are related to
the solutions of linearized equations in GR with the same $T^{\mu}_{\nu}$
in the follwing way: if $g^{gr}_{\mu\nu}(G,x)$ is a known solution of the
Einstein's equation in the weak field approximation for a given
$T^{\mu}_{\nu}$, then the BD solution corresponding to the same
$T^{\mu}_{\nu}$, in the weak field approximation, is given by 
$$
g^{bd}_{\mu\nu}(x) = [1-G_{0}\epsilon(x)]g^{gr}_{\mu\nu}(G_{0},x)
\eqno{(3.2)}
$$
where $\epsilon(x)$ must satisfy
$$
\Box\epsilon(x)= {8\pi T\over{(2\omega+3)}},
\eqno{(3.3)}
$$
and $G$ is replaced by $G_{0}$ defined earlier in this section.
Hence, to get the spacetime for vacuumless global and gauge string one has
to solve the equation (3.3) with $T^{\mu}_{\nu}$ given in section 2.
\vspace{5mm}
\subsection{Global String}
Using equations (2.4) and (2.5) with $\alpha = 0$, one can calculate $T =
T^{\mu}_{\nu}$ in equation (3.3) as
$$
T = 2f^{'2} +
{f^{2}\over{C}} + 4V(f).
\eqno{(3.4)}
$$ 
Now under weak field approximation $C = r^{2}(1+\gamma(r))$ where
$\gamma(r)<<1$. Hence  (3.4) becomes
$$
T = 2f^{'2} +
{f^{2}\over{r^{2}}} + 4V(f)
$$
which on substitution of (1.2) and (2.1) with $r>>\delta$ becomes
$$
T=Ar^{-2n\over{(n+2)}}
\eqno{(3.5)}
$$
where $A = a^{2}M^{2}({1\over{\delta}})^{4/(n+2)}[{8\over{(n+2)^{2}}} + 1
+
{4\over{a^{n+2}}}]$. Putting (3.5) in
(3.3)
and using (2.4) and (2.5d), we get
$$ 
\epsilon^{''} + {2\epsilon^{'}\over{r}} = {8\pi A\over{2\omega+3}}r^{-2n\over{(n+2)}}
$$
which on integration yields
$$
\epsilon = ({8\pi\over{2\omega+3}}) C({r\over{\delta}})^{4\over{n+2}} -
{D\over{2\omega+3}}({r\over{\delta}})^{-3},
\eqno{(3.6)}
$$
where $D$ is an arbitary integration constant and $C= a^{2}M^{2}[{8\over{(n+2)^{2}}} +
1 +
{4\over{a^{n+2}}}]{(n+2)^{2}\over{4(n+6)}}$.
Hence the complete line element for a global vacuumless string in BD
theory under weak field approximation for $r>>\delta$ is given by
$$
ds^{2} = [1-G_{0}\epsilon(r)]ds^{2}_{cv}(G_{0})
\eqno{(3.7)}
$$
where $\epsilon(r)$ is given in equation (3.6) and $ds^{2}_{cv}(G_{0})$
is the corrsponding line element obtained by Cho and Vilenkin ~\cite{CV2}
in GR with $G$ is replaced by $G_{0}$ defined earlier in this section. One
should keep in mind that $r$ is bounded by equation (2.1a) for the form of
$f(r)$ given in equation (2.1) to be valid.

To have an idea of the motion of the particles, one can calculate the
radial
acceleration vector $\dot{v}^{1}$ of a particle that remains stationary (i.e.,
$v^{1}=v^{2}=v^{3}=0$) in the field of the string. Now,
${\dot{v}}^{1}=v^{1}_{;0}v^{0}=v^{0}\Gamma^{1}_{00}v^{0}$. Hence using the
line
element (3.7) one can calculate ${\dot{v}}
^{1}$ which becomes
$$
{\dot{v}}^{1} =
{1\over{2}}(1-G_{0}\epsilon)^{-2}(1+2\Phi)^{-1}[2\Phi^{'}(1-G_{0}\epsilon)-
G_{0}(1+2\Phi)\epsilon^{'}].
\eqno{(3.8a)}
$$
Now as $G_{0}\epsilon<<1$ and also $\Phi<<1$ for the linearized approximation to be
valid one can approximate  equation (3.8) to write
$$
{\dot{v}}^{1} = \Phi^{'} - {G_{0}\over{2}}\epsilon^{'}.
\eqno{(3.8b)}
$$
For $\epsilon=constant$ that is in GR the acceleration vector is always -ve and the
gravitational force is repulsive. But for $\epsilon^{'}\neq 0$ one can check that 
${\dot{v}}^{1}$ is -ve or +ve depending on the arbitary constant $D$. For example, for
$n=2$ in GR the repulsive force is independent of radial distance. But in this case for
$n=2$
$$
{\dot{v}}^{1}= -{KG_{0}M^{2}\over{\delta}}-{4\pi
cG_{0}\over{(2\omega+3)\delta}}-
{3DG_{0}\over{2\delta(2\omega+3)}}({r\over{\delta}})^{-4},
$$
and in this case one can see that the gravitational force varies with radial
 distance
and with -ve $D$ one can have a attractive gravitational force as well  
, as $r$ is bounded by equation (2.1a) for the linearized approximation as
mentioned
in section 2.
So in BD theory the vacuumless global string can have both repulsive and
attractive gravitational effect.
 \subsection{Gauge String}

For gauge string using equation (2.8), equation (3.3) becomes
$$
\epsilon^{''}+{2\epsilon^{'}\over{r}} = {a^{2}\over{r^{2}(2\omega+3)}}
$$
which on integration yields
$$
\epsilon =
{a^{2}\over{(2\omega+3)}}Ln(r/\delta)-{P\over{(2\omega+3)r}}
\eqno{(3.9)}
$$
where $P$ is an arbitary integration constant and
we have put another integration constant equal to
${-a^{2}Ln{\delta}\over{(2\omega+3)}}$. Hence
the metric
for a gauge vacuumless string in BD theory under weak field approximation can
be written according to equation (3.2) as
$$
ds^{2} = [1-G_{0}\epsilon]ds_{cv}^{2}(G_{0})
\eqno{(3.91)}
$$
where $ds_{cv}^{2}(G_{0})$ is the metric given in equation (2.9) with $G$
is
replaced by $G_{0}$. Here also $r$ is bounded by equation (2.8a). 
One can calculate the acceleration vector
${\dot{v}}^{1}$ for a particle remaining stationary with respect to the
string. Assuming $|\epsilon G_{0}|<<1$ and $\Phi<<1$ one can again
approximate the acceleration vector as 
$$
{\dot{v}}^{1} = \Phi^{'} - {G_{0}\over{2}}\epsilon^{'},
$$
Which becomes
$$
{\dot{v}}^{1}={G_{0}\over{r^{2}}}[2\pi
a^{2}r-{a^{2}r\over{2(2\omega+3)}}-{P\over{2(2\omega+3)}}].
\eqno{(3.92)}
$$
Now as $r$ is bounded by (2.8a) for the form of the energy momentum tensor
(2.8) to be
 valid one can
have repulsive as well as attractive gravitational effect for $-ve P$ and $+ve P$
respectively.
 But this is not so in GR as for $\omega\rightarrow\infty$ $\epsilon^{'}$
vanishes.
 \section{Conclusion}

Recently there have been claims that the universe must possess a not yet
identified component usually called {\it quintessence} matter or Q matter,
besides its normal content of matter and radiation. These claims have been
prompted at the realization that the clustered matter component can be at
most one third of the critical density. That is why there must be some
additional nonclustered component if the critical density predicted by the
inflationary models is to be achieved. Examples of Q matter are
fundamental fields or macroscopic objects and network of vacuumless
strings may be one such good examples as scalar field with potential like
(1.2) can act as quintessence models~\cite{VIL}.
In this paper we have examined the gravitational field of
vacuumless strings in the BD
theory under the weak field approximation of the field equations. In doing so,
we have followed the method of Barros and Romero ~\cite{BR} which has been
suggested recently. For both global and gauge vacuumless strings, the
spacetimes are conformally related to that obtained earlier by Cho and
Vilenkin in GR~\cite{CV2}. Both the spacetimes reduces to the corresponding
GR solutions for $\omega\rightarrow\infty$ limit.  It has been shown that
both the global and gauge string can have attractive as well as repulsive
gravitational effect on a test particle freely moving in its spacetime
which is not so in GR where the global string has only repulsive and gauge
string has the attractive gravitational effect.
 As the trajectories of
the light rays, which are given by the null
geodesics, the only change involved in BD theory is the replacement of $G$ by
an new $\omega$ dependent "effective" gravitational constant $G_{0}=
({(2\omega+3)\over{(2\omega+4)}})G$ and for $\omega$ to be consistent with 
solar system experiment and observation, $\omega \sim 500$~\cite{Will}, this means
that photons travelling in the spacetime will experience a decrease of gravitational
constant as $G_{0}\sim 0.999G$.
Therefore, it follows that the distortion
of the isotropy of the CMBR due to the gravitational field of the vacuumless
strings in BD theory may be calculated directly from the results obtained in
GR. A detail analysis of the full nonlinear Einstein's equations will
certainly give more insight to the problem and for that a detail numerical
calculation should be done which will be the aim of our future study.

\end{document}